\documentclass[aps,prb,10pt,twocolumn]{revtex4-2}

\usepackage{amsmath} 
\usepackage{amssymb}
\usepackage{mathrsfs}
\usepackage{amsbsy}
\usepackage{bm}
\usepackage{graphicx}
\usepackage{times}
\usepackage{array}
\usepackage{color}
\usepackage{upgreek}
\usepackage[dvipsnames]{xcolor}
\usepackage{dsfont}
\usepackage{ulem}
\usepackage{textpos}
\usepackage{subfigure}
\usepackage{braket}
\usepackage[export]{adjustbox}\usepackage[colorlinks=true,citecolor=NavyBlue,linkcolor=RubineRed,urlcolor=Cerulean,pdfencoding=auto]{hyperref}

\newcommand{\eref}[1]{Eq.~\eqref{#1}}
\newcommand{\fref}[1]{Fig.~\ref{#1}}
\newcommand{\hc}{^\dagger}

\DeclareMathOperator{\da}{\downarrow}
\DeclareMathOperator{\ua}{\uparrow}

\begin{document}
\title{Fidelity of photon-mediated entanglement between remote nuclear-spin multi-qubit registers}

\author{W.-R.~Hannes}
\author{Regina~Finsterhoelzl}
\author{Guido~Burkard}
\affiliation{Department of Physics, University of Konstanz, 78457 Konstanz, Germany }


\begin{abstract}
	The electron spin of a nitrogen-vacancy center in diamond lends itself to the control of proximal $^{13}$C nuclear spins via dynamical decoupling methods, possibly combined with radio-frequency driving. Long-lived single-qubit states and high-fidelity electron-nuclear gates required for the realization of a multiqubit register have already been demonstrated. Towards the goal of a scalable architecture, linking multiple such registers in a photonic network represents an important step. Multiple pairs of remotely entangled qubits can enable advanced algorithms or error correction protocols.  We investigate how a photonic architecture can be extended from the intrinsic nitrogen spin to multiple $^{13}$C spins per node. Applying decoherence-protected gates sequentially, we simulate the fidelity of creating multiple pairs of remotely entangled qubits. Even though the currently achieved degree of control of $^{13}$C spins might not be sufficient for large-scale devices, the two schemes are compatible in principle. One requirement is the correction of unconditional phases acquired by unaddressed nuclear spins during a decoupling sequence. 
\end{abstract}

\maketitle	

\section{Introduction}

Electron and nuclear spins associated with the negatively charged nitrogen-vacancy (NV) center in diamond are considered as a promising platform for several quantum technologies, including quantum sensing and quantum computing \cite{Awschalom2018,Atature2018,Wolfowicz2021,Ruf2021}. The benefit of such a hybrid quantum register is that the electron spin can be manipulated fast and with high-fidelity, 
while the more robust nuclear spins may serve as long-term quantum memory nodes. 
A major challenge on the way to scalable quantum networks is the creation of long-range entanglement links between different nodes. While two nearby NV centers within one crystal can be directly coupled via magnetic dipole-dipole interaction \cite{Neumann2010}, a more flexible solution is to use spatially separate nodes and to establish the link between them via a photonic network along with spin-photon interfaces~\cite{Kalb2017}.

As an example for such a system, the three-level $\Lambda$ system consisting of the two $\ket{\pm 1}$ ground states and one optically excited state allows the creation of spin-photon entanglement by optical absorption and emission \cite{Togan2010}. Then the coupling of remote nodes is achieved by linear-optical elements, in particular by interfering the emitted photons from a pair of nodes at a beam splitter and post-selecting the entangled state of the two qubits \cite{Barrett2005}. 
Such a scheme has been experimentally realized a decade ago \cite{Bernien2013}, and has recently enabled 
a three-node quantum network \cite{Pompili2021}. 
However, scaling up these networks further is challenging due to the low spin-photon coupling efficiency, 
and because embedding NV centers in nanocavities comes with drawbacks. 
In this respect, silicon-vacancy (SiV) color centers are more suitable for the integration of nodes within an optical fiber network.
Strong cavity coupling has enabled high-contrast spin-dependent reflection from SiV-cavity systems, leading to a very  efficient (nearly deterministic) nanophotonic interface \cite{Nguyen2019,Nguyen2019a}. Combined with a controllable nuclear spin such as the $^{29}$Si spin, this has led to a two-qubit register with nuclear spin–photon gates \cite{Stas2022}.
Recently such a system has been integrated with telecom-band systems to demonstrate the viability for scalable quantum networking applications \cite{Bersin2023}.
On the other hand, so far multiqubit control and large memory nodes have been realized mostly by means of the NV center with surrounding $^{13}$C nuclear spins, which thus remains a very promising platform for such applications at the moment. 

In this paper, we focus on the theoretical proposal by Nemoto \textit{et al.} \cite{Nemoto2014} for a photonic architecture based on a simple module comprising an optical cavity containing a single
NV center, and discuss its extension to a larger number of nuclear spin memory qubits to realize high-fidelity remote quantum gates and potentially advanced error correction protocols. 
In recent years, the coherent and universal control over the intrinsic nitrogen \cite{Sar2012} and surrounding $^{13}$C nuclear spins has been steadily improved by means of dynamical decoupling (DD) methods \cite{Taminiau2012,Taminiau2014,Abobeih2018,Bradley2019}.
A combination of DD-based control \cite{Taminiau2012} with a single-photon-based entangling protocol has already been used to implement an entanglement distillation protocol between distant nodes~\cite{Kalb2017} based on one nuclear spin pair.
A more recent DD technique by Bradley \textit{et al.} additionally employs radio frequency (RF) driving, resulting in high-fidelity control even for  weakly coupled $^{13}$C spins~\cite{Bradley2019} and thus in a larger number of potential memory qubits. 
This method, in short DDRF, will be analyzed in more detail to assess the compatibility with photonic networks as considered by Nemoto \textit{et al.}~\cite{Nemoto2014}. 

Our paper is organized as follows. 
In Sec.~\ref{sec:model} we discuss some practical issues arising when trying to combine the photonic-network based entanglement scheme (Nemoto scheme) with a DD/DDRF-based nuclear spin quantum register (Bradley scheme). 
In Sec.~\ref{sec:results} we assess the fidelity of the entanglement transfer from the remote electrons onto the remote nuclei.
We build our model successively by first excluding all other spins except one target nuclear spin, and then including either additional bath spins or additional spins belonging to the quantum register. 
A total gate fidelity is obtained by multiplying with the fidelity of the electron-electron entanglement scheme, which is not modeled explicitly in this study, and potentiating with the number of qubit pairs. In Sec.~\ref{sec:discussion} we list further error sources which are not included by our model, and discuss the usefulness of the scheme with regard to scalable quantum networks. We summarize our findings in Sec.~\ref{sec:conclusions}.

\section{Remote nuclear spin entangling scheme}
\label{sec:model}

\subsection{Generation of a single pair of remote entangled qubits}

We consider two remote NV centers (labeled $a$ and $b$), each with $p$ nuclear spins $n_i^{a,b}$ ($i=1,\dots,p$), which we aim to entangle pairwise. 
Since only one qubit, realized by the electron spin, is available for the interaction with the cavity photons, a multiplexing approach~\cite{LoPiparo2019} is not applicable and entangled pairs can be created only sequentially.
The generation of entanglement between one pair of remote nuclear spins can be divided into three steps, each consisting of several gates (see \fref{fig:pseudocircuit}): (i) initialization of the electron and nuclear spin in each node, (ii) photon-mediated entanglement generation between remote electron spins and concomitant decoupling of nuclear spins, and (iii) transfer of the entanglement from the electron spins onto the nuclear spins, followed by a measurement of the electron spin as entanglement is created non-deterministically. Here we discuss these steps in more detail:
\begin{figure}
	\centering
	\includegraphics[width=0.9\linewidth]{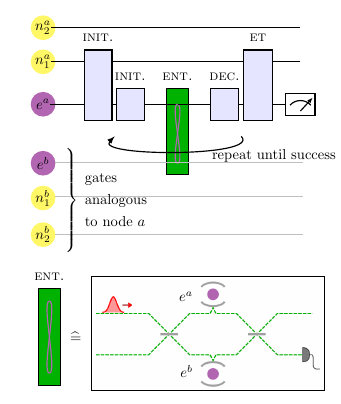}
	\caption{Circuit diagram for the entanglement of one nuclear spin ($n_1^a$) of node $a$ with the corresponding nuclear spin ($n_1^b$) in another node $b$, via the electron spins $e^a$ and $e^b$ of nodes $a$ and $b$. The sequence of gate blocks consists of nuclear spin initialization (INIT.), electron spin initialization (INIT.), non-deterministic remote electron-spin entanglement creation (ENT.), decoupling sequence (DEC.), and entanglement transfer to the nuclear spin (ET). The process is repeated subsequently for nuclear spins $n_2^a$ and $n_2^b$ (see \fref{fig:circuit2nv2n}). Bottom panel: schematic drawing of a single photon–based Michelson interferometer which can be used for entangling the remote electron spins. Each node is embedded in an optical cavity with spin-dependent reflection coefficient. The interaction of the photon with the cavities and with beam splitters (shown in gray) leads to non-deterministic entanglement creation when the detector clicks at the dark port~\cite{Nemoto2014}.}
	\label{fig:pseudocircuit}
\end{figure}

(i) In both Nemoto \textit{et al.}\ \cite{Nemoto2014} and Bradley \textit{et al.}\ \cite{Bradley2019} the initialization of the electron spin is based on a projective measurement via spin-photon entanglement. The details differ, because the setup by Nemoto \textit{et al.}\ does not allow resonant excitation of a single transition or fluorescence detection \cite{Robledo2011}, which are used in Bradley \textit{et al.} We therefore choose the quantum non-demolition measurement (QND) approach by Nemoto \textit{et al.}, where the electron spin is probed by a single photon a number of times. 
The initialization of the nuclear spin, on the other hand, can be achieved by a combination of single-qubit and two-qubit gates. The single-qubit gates, i.e. electron-spin rotations, are performed by microwave (MW) driving fields. The only difference is thus in the two-qubit gate: In the Nemoto scheme, it is performed by the natural hyperfine generated CPHASE gate. 
The Bradley scheme swaps the state of the initialized electron spin onto a particular $^{13}$C spin by using two CNOT gates (see Fig.~4(a) in \cite{Bradley2019}). In order to keep other nuclear spins decoupled, this CNOT gate is realized by a conditional DDRF sequence. 
It is worth noting that as a tradeoff to the large number of available qubits the control of weakly coupled $^{13}$C spins is  roughly three orders of magnitude slower than controlling the intrinsic nuclear spin. 
(For the case of the $^{14}$N spin, initialization is performed by a measurement-based scheme which heralds the preparation in a particular eigenstate \cite{Bradley2019}.)

(ii) The remote entanglement is created in the Nemoto protocol by single-photon interference using beam splitters, combined with electron spin-dependent reflection from the cavity. We can apply this scheme to the Bradley setup. The successful event is a photon detector click at the dark port. In the absence of losses its probability is 0.125, so that a repeat-until-success scheme is required (but alternative schemes with a higher probability exist). In the application to the system considered here, comprising many $^{13}$C nuclear spins, an important point is to keep the nuclear spins decoupled from the electron spin during this process. In the Nemoto protocol, after each attempt, whether successful or not, a spin-echo technique is applied to decouple the $^{15}$N spin. There is, however, a slight phase rotation, which needs to be corrected later. Due to the decoupling, there is no need to reinitialize the $^{15}$N spin in the case that the attempt has failed (see arrow in \fref{fig:pseudocircuit}). 
For the Bradley setup, most nuclear spins are coupled weakly, so that their 
natural coherence time might be long compared to the duration of the photonic scheme (including repetitions). 
In the presence of strongly coupled nuclear spins, DD can be applied as soon as the electron is initialized to the $\ket{+}$ state prior to the spin-photon interaction. Note that after the detection of a photon at the dark port, the electron state is $\ket{0}_{a}\ket{1}_{b}-\ket{1}_{a}\ket{0}_{b}$, which (like $\ket{+}$) is also left invariant by DD $\pi$-pulses applied simultaneously in both nodes. 
Since the intrinsic $^{14}$N spin is strongly coupled it requires DD on its own, unless it is in a polarized state (i.e., not storing a memory qubit), where it remains stable due to the large quadrupole splitting.
We thus conclude that all nuclear spins can be decoupled during the creation of remote node entanglement. 
However, analogous to the slight phase rotation mentioned above for the Nemoto scheme, each nuclear spin is subject to an unconditional rotation, depending on its hyperfine parameters. This needs to be fixed if the nuclear spin stores a memory qubit. 

(iii) Transferring the remote entanglement requires the same type of control as initializing nuclear spins. While in Nemoto the natural CPHASE gate is used, this can again be replaced by the CNOT gate obtained from the DDRF approach. Bradley reported the control of both $^{14}$N or $^{13}$C nuclear spins with DDRF \cite{Bradley2019}, so no separate technique is required. Some $^{13}$C spins may be controllable with DD alone depending on their hyperfine tensor. In addition to the two-qubit gate, single-qubit gates may be needed to bring  the  entangled state into the desired form. The entanglement transfer is completed by measuring the electron spins, which projects the pair of nuclear spins into one of four possible maximally entangled states. 

\begin{figure}
	\centering
	\includegraphics[width=1\linewidth]{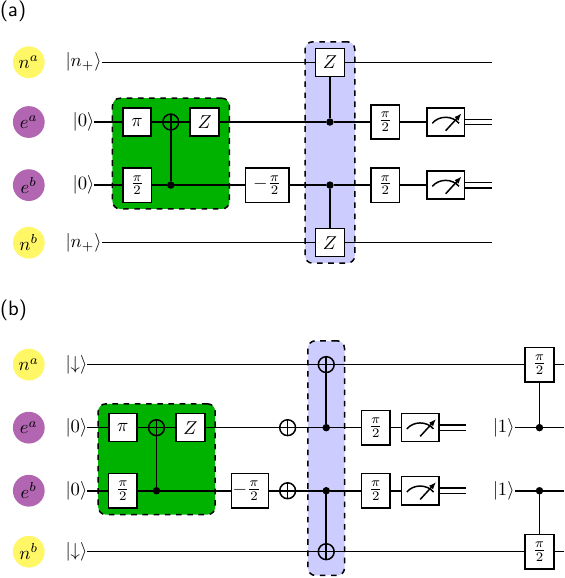}
	\caption{(a) Effective circuit diagram for the Nemoto scheme, (b) CNOT-based alternative resulting in the same set of nuclear states. The gates in the green box effectively represent the entangling scheme via the photonic network (in the successful case); directly afterwards the electron-electron state is $\ket{0^a1^b}-\ket{1^a0^b}$. Note that single-qubit gates depend on the electron spin being in the state $|1\rangle$, as seen at the end of the circuit. Angles denote rotations about the y axis.
 Labels and colors of nuclear and electron spins are as in \fref{fig:pseudocircuit}.}
	\label{fig:circuitnemoto}
\end{figure}

The circuit diagram in \fref{fig:circuitnemoto}~(a) effectively shows the generation of remote nuclear entanglement in the Nemoto protocol, omitting the initialization and decoupling steps, timing issues, and repeat-until-success scheme. After the photon-based entangling scheme, effectively represented by the gate group in the green box, the electron pair is in the singlet (Bell) state $(\ket{01}-\ket{10})/\sqrt{2}$. The electron spin in node $b$ is then rotated by $-\pi/2$ around Y to transform this state into a two-qubit cluster state $(\ket{0+}-\ket{1-})/\sqrt{2}$. The following CZ operations are obtained naturally by waiting half a period of the effective hyperfine interaction 
$A_\text{net} = A_\parallel - \frac{A_\perp^2}{2\Gamma}$ with $\Gamma = D + \gamma_\mathrm{e} B_z - \gamma_\mathrm{n} B_z$ \footnote{Note the differing sign in $A_\text{net}$ in \cite{Nemoto2014}}. The Hamiltonian of the NV-$^{15}$N system is
\begin{equation}\label{eq:HamNV14N}
H = D S_{z}^{2} + \gamma_{e} B_z S_{z} -\gamma_{n} B_z I_{z} + A_{\parallel} S_{z} I_{z} + A_{\bot} (S_{x}I_{x}+S_{y}I_{y}),
\end{equation}
where $D$ is the zero field splitting of the electron
triplet state, and the operators $S_{\alpha}$ and $I_{\alpha}$ are the spin-1 and spin-1/2 operators of the electron and nuclear spins respectively.
The magnetic field $B_z$ is applied along the NV axis, and $\gamma_{e}$ ($\gamma_{n}$) is the gyromagnetic ratio of the electron (nuclear) spin.
$A_{\parallel}$ ($A_{\bot}$) denotes the axial $z$ ($x$ and $y$) component of the hyperfine tensor. 
After a further single-qubit $\pi/2$ rotation, the electron spins are measured in the computational basis so that the nuclear spins are projected onto a two-qubit cluster state. Up to measurement-dependent Pauli corrections, which are classically tracked, the combined nuclear state is $\tfrac{1}{\sqrt{2}} (\ket{n_+\ua}+\ket{n_-\da})$, where $\ket{n_\pm}=\tfrac{1}{\sqrt{2}} (\ket{\da}\pm\ket{\ua})$.


\fref{fig:circuitnemoto}(b) shows a similar circuit resulting in the same set of possible outcome states for the four different measurement results. Here the central two-qubit gate in the entanglement transfer process is a CNOT instead of the CZ. The additional two-qubit gate after the measurement is optional; it is included here to obtain the same nuclear qubit states as from \fref{fig:circuitnemoto}~(a). Note that this scheme requires initialization of nuclear spins in a polarized state only, while (a) requires the superposition state $\ket{n_+}$ prior to electron-electron entanglement. This faster initialization compensates the slower CNOT entangling gate so that the total durations of the two circuits are expected to be comparable.


\subsection{Extension to multiple pairs of remote entangled qubits}

\begin{figure}[t!]
	\centering
	\includegraphics[width=0.9\linewidth,valign=t]{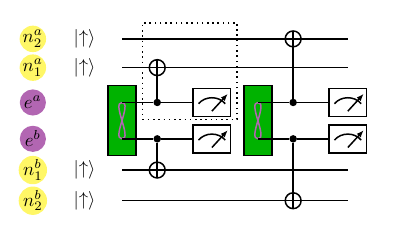}
	\caption{Schematic circuit diagram for creating remote (node-to-node) entanglement in two pairs of qubits. Electron initialization and single-qubit gates have been omitted. 
	The transfer onto nuclear spins is indicated by CNOT gates and electron measurements. 
 The dotted line frames the elementary process which is applied multiple times. 
 Each node can comprise several additional bath nuclear spins 
 which are not shown here.
	}
	\label{fig:circuit2nv2n}
\end{figure}

In the extension of this scheme to multiple pairs of entangled nuclear spins several possible problems need to be taken care of. 
The most natural way seems to be the sequential application of the scheme outlined above to each addressed nuclear spin. 
This would correspond to initializing each nuclear spin directly before addressing it, but this might not be the best choice because of the deleterious effect of nuclear-spin preparation on other spins. As an application of their DDRF scheme, Bradley \textit{et al.}\ report the preparation of a GHZ state (Fig.~5(a) in \cite{Bradley2019}), in which case they initialize all nuclear spins first. With this choice, the corresponding scheme to create two pairs of entangled qubits is shown in simplified form in \fref{fig:circuit2nv2n}. 
While entangling one nuclear pair, it is obvious that other nuclear spins need to be protected in their state, whether it stores remote entanglement or not yet. In a simplified manner, we can thus investigate the ideal elementary process enclosed by the dotted line, which is sequentially applied to each target pair. 
If we assign the fidelity ${F}_{en\bar{n}}$ to such an elementary electron to nuclear storage process, 
where $\bar{n}$ stands for $p-1$ unaddressed nuclear spins, 
the total gate fidelity for entangling $p$ pairs of nuclear spins (e.g., $p=2$ in \fref{fig:circuit2nv2n}) in the case of uncorrelated errors amounts to
${F} = {F}_{ee}^{p} {F}_{en\bar{n},a}^{p} {F}_{en\bar{n},b}^{p} $, where $F_{en\bar n,i}$ denotes the electron-nuclear transfer fidelity in node $i$.
If we take the two nodes as identical for simplicity, then ${F} = {F}_{ee}^{p} {F}_{en\bar{n}}^{2p}$.
As expected from previous studies~\cite{Bradley2019,Takou2023}, we find that the presence of further bath spins, which are never targeted, further diminishes the total gate fidelity due to the unavoidable imperfection of the DD technique. 

The key process that is needed thus consists of a controlled gate on a target nuclear spin qubit, which leaves all other nuclear spins untouched. 
In general the gate fidelity of this process will depend on hyperfine parameters and on the physical implementation of the gate. We calculate a gate fidelity for the elementary process with the CNOT gate realized by a DDRF sequence in the next Section.
In this study we do not explicitly model the electron entangling scheme, but fix the value ${F}_{ee}$ to 0.99 \cite[]{Nemoto2014}.


\section{DDRF simulations}
\label{sec:results}


A single node consists of the electron spin, the nuclear spins belonging to the multiqubit register, and possibly further nuclear spins forming a bath. In general this is a highly complex system, but since we employ the secular approximation and neglect direct interactions between nuclear spins (see below), the dynamics of different nuclei are independent. Therefore, a single simulation needs to include only a single $^{13}$C nuclear spin coupled to the NV center electron spin by the hyperfine interaction. Depending on whether the hyperfine parameters are fitting to the DDRF sequence parameters or not, the modeled spin is a target or an unaddressed spin of the controlled operation. 
The obtained evolution operators from different simulations can be combined to give the gate fidelity for more complex processes.
In general, such a gate fidelity can be calculated according to~\cite{Pedersen2007,Ghosh2010},
\begin{equation}\label{eq:gatefidelity}
	{F} = \frac{d + \left| \mathrm{Tr} \left[ V_\text{ideal}\hc V_\text{actual} \right] \right|^2}{d(d+1)},
\end{equation}
where $V_\text{ideal}$ and $V_\text{actual}$ are the unitaries describing the desired (ideal) and actually obtained quantum operations in a Hilbert space of dimension $d$ (e.g., $d=8$ for the elementary process highlighted in \fref{fig:circuit2nv2n}). 
By applying this formula to the evolution within one node, an overall phase change of the state in one node is not captured, but this is irrelevant.
We apply the method described above to the simplest system with only one target spin present in Sec.~\ref{ssec:targetspin}, and with one additional unaddressed spin present in Sec.~\ref{ssec:unaddrspin}, which are combined in Sec.~\ref{ssec:gatefid}.
Another type of spin, which is not part of the quantum register (i.e., never used as target spin) is referred to as  bath spin and its influence on the fidelity is analyzed separately as described in Sec.~\ref{ssec:bathspin}. 

\begin{figure}
	\centering
	\includegraphics[width=\linewidth]{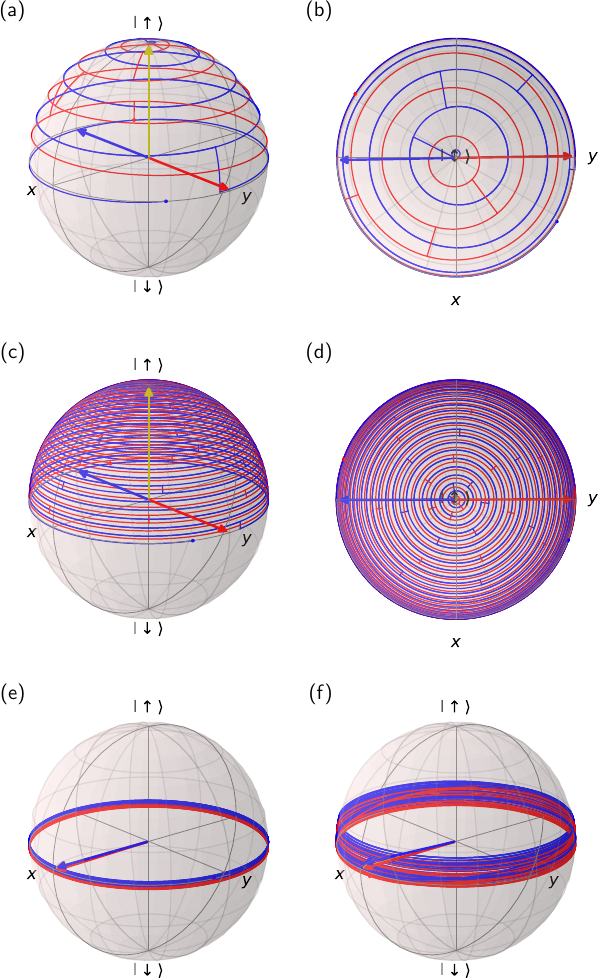}
	\caption{Bloch spheres showing the nuclear spin trajectories of target (a-d) or unaddressed (e-f) spins during a DDRF CROT gate. (b) and (d) are top views of (a) and (c), respectively. The initial state, shown by the yellow arrow, is chosen as $\ket{\ua}$ ($\ket{n_+}$) for the addressed (unaddressed) nuclear spins. The blue (red) path shows the evolution when the electron is initially in the state $\ket{0}$ ($\ket{1}$). After the drawn evolution ends (marked by dots), another change of rotating frame, $R_{\sigma}(2N\tau)\hc$ from \eref{eq:totalevogeneral}, and the phase correction $R_z(-N \omega_1 \tau)$ are applied to arrive at the final state (in the rotating frame of \eref{eq:Hgeneral}) indicated by the blue/red arrows.
    In (a-b) we set $N=8$ for a clearer image, while in (c-f) our default value $N=48$ is used. The hyperfine parameters are ${A_\parallel} = 2\pi \cdot 50\ \text{kHz}$ and ${\beta}=0$ for the target spin. For the unaddressed spins, $\bar{A_\parallel} = 2\pi \cdot 30\ \text{kHz}$ in both (e-f), and they differ by (e) $\bar{\beta}=0$ and (f) $\bar{\beta}=0.1$. In (f), with $\bar{\beta}\neq 0$ the evolution is shown in alternating rotating frames $R_0(t)$ and $R_1(t)$, see App.~\ref{sec:DDRF_equations}.
    }
	\label{fig:Blochsphere}
\end{figure}

In the choice of the electron qubit subspace (states \(\ket{0}\) and \(\ket{1}\)) we follow Ref.~\cite{Bradley2019}, where these qubit computational basis states correspond to the \(m_s = 0\) and \(-1\) states with \(m_s\) being the magnetic quantum number, while \(m_s = 0\) and \(+1\) are used in the Nemoto protocol.
Both the DDRF scheme and the Nemoto protocol require a constant magnetic field applied along the NV axis (\(\hat{z}\)). Nemoto \textit{et al.}~\cite[]{Nemoto2014} suggest 20~mT to separate the \(m_s = \pm 1\) electron states. The accuracy of the DDRF scheme benefits from a stronger field; 40.3~mT are chosen by Bradley \textit{et al.}~\cite{Bradley2019}, which is still compatible with the Nemoto protocol  and thus adopted here. 

The Hamiltonian of the NV-$^{13}$C system is given by
\begin{equation}\label{eq:Hgeneral}
H = \omega_{L} I_{z} + A_{\parallel} S_{z} I_{z} + A_{\bot} S_{z} I_{x} + 2\Omega\cos(\omega t + \phi)I_x,
\end{equation}
where the electron energy splitting (i.e., the terms $D S_{z}^{2} + \gamma_{e} B_z S_{z}$ from \eref{eq:HamNV14N}) has been removed by switching to a rotating frame, and the non-secular terms have been neglected.
The operator $I_{\alpha}$ is again a spin-1/2 operator, now for the $^{13}$C nuclear spin.
The magnetic field $B_z$ applied along the NV axis enters through the nuclear spin Larmor frequency $\omega_L = \gamma_{n} B_z$, where $\gamma_{n}$ is the gyromagnetic ratio of the $^{13}$C spin.
Furthermore, $A_\parallel$ ($A_\perp$) denotes the component of the electron-nuclear hyperfine interaction parallel (perpendicular) to $\hat{z}$, where the presence of only one nuclear spin allows the choice of $\hat{x}$ for the perpendicular direction. Note that $A_\perp$ is defined differently from $A_\bot$ in \eref{eq:HamNV14N}.
The final term describes the RF driving of the nuclear spin by a field polarized along $x$ with frequency $\omega$, phase $\phi$, and Rabi frequency $\Omega$.   

Following Ref.~\cite{Bradley2019}, the DD sequence is chosen as ($\tau$ - $\pi$ - $2\tau$ - $\pi$ - $\tau$)$^{N/2}$, where $\pi$ is a $\pi$-pulse on the electron spin, $2\tau$ is the interpulse delay, and $N$ is the total number of electron decoupling pulses \cite{Kolkowitz2012, Taminiau2012, Zhao2012}. 
The evolution of the nuclear spin during this sequence is analyzed separately for the two initial electron eigenstates: $\ket{0}$ and $\ket{1}$.
When the electron is in the state $\ket{0}$, the nuclear spin precesses with the bare Larmor frequency $\omega_L$ around \(\hat{z}\).  For the electron state $\ket{1}$, its frequency is $\omega_{1} = \sqrt{A_{\perp}^{2}+(\omega_{L}-A_{\parallel})^{2}}$, and the quantization axis $A_{\perp} \hat{x} + (\omega_{L}-A_{\parallel}) \hat{z}$ is tilted by an angle $\beta$ from $\hat{z}$, defined by $\cos(\beta) = (\omega_L - A_\parallel)/\omega_{1}$.
For a conditional rotation, the RF pulse phases are chosen as~\cite{Bradley2019} 
\begin{equation}\label{eq:rfphases}
    \phi_k = \left\{ 
    \begin{array}{ll} 
    (k-1)\phi_\tau + \varphi + \pi, & \quad k\text{ odd}, \\
    (k-1)\phi_\tau + \varphi, & \quad k\text{ even},
    \end{array}
    \right.
\end{equation}
where $\phi_\tau = (\omega_L-\omega_1)\tau$ and $k = 1,\dots,N+1$ labels each successive RF pulse.
The rotation axis of the gate is adjusted by the phase $\varphi$.
Since we approximate the electron $\pi$-pulses as instantaneous, a piecewise treatment renders the simulation effectively time-independent. 
The derivation of the evolution under the DDRF sequence closely follows Ref.~\cite{Bradley2019} and is given in App.~\ref*{sec:DDRF_equations}. Its solution is performed using the QuTip Python package \cite{Johansson2013}.

Since we take the quantum register of Ref.~\cite{Bradley2019} as a prototype, we adopt some of the experimental parameter values.
In particular, we set the Larmor frequency to \(\omega_L = 2\pi \cdot 432\ \text{kHz}\) (corresponding to a $B$ field of about 40~mT), the number of \(\pi \) pulses to \(N=48\) (except in \fref{fig:Blochsphere}~(a)), and the interpulse duration as an integer multiple of the Larmor precession time \(\tau_L = 2\pi/\omega_L\), namely \(\tau = 8 \tau_L\). 
The parallel hyperfine components of the $^{13}$C nuclei are in the kHz range; our primary target nucleus is chosen with \(A_\parallel = 2\pi \cdot 50\ \text{kHz}\).
For the considered magnetic field and the entire set of $^{13}$C spins in Ref.~\cite{Bradley2019}, the tilting angle $\beta$ varies in the range from 0 to roughly 0.1. The target spins can be usually selected such that $\beta\ll 0.1$.

The remaining parameters are fixed by the desired electron-nuclear gate for storage of the electron spin qubit into the addressed nuclear spin. 
Among the available nuclear spins, the target spin is selected by the condition \(\omega=\omega_1\). 
Since our desired gate is a maximally entangling CROT gate, i.e. controlled-\(R_x(\pm\tfrac{\pi}{2})\), the driving Rabi frequency $\Omega$ is chosen to give a \(\pi/2\) rotation over one sequence. In the simplified DDRF model from Ref.~\cite{Bradley2019} (neglecting \(A_\perp\) and off-resonant driving) this would correspond to \(N \Omega \tau = \tfrac{\pi}{2}\), while in the more accurate model used here this needs to be corrected by a  factor \(\lesssim 1\). The optimal value of this factor is determined via fitting. The phase offset \(\varphi\) can be freely chosen such that the target spin ends up in the desired state. Here, we can set \(\varphi=0\) by making use of a different phase correction described in the following paragraph. 

An inherent effect of the DD technique is an unconditional rotation of all other nuclear spins depending on their hyperfine parameters~\footnote{This unconditional rotation could be avoided by using a symmetrized sequence, which would symmetrically involve both $m_s=\pm 1$ states and opposite magnetic field amplitudes $\pm B_z$.}. 
Due to the relatively strong magnetic field, this rotation is roughly around the $z$ axis, c.f.~\fref{fig:Blochsphere} (e-f), and the total azimuthal phase acquired by each spin can be approximated by $N (\omega_L+\omega_1) \tau$, depending on its individual parameter $\omega_1$. 
This individual precession necessitates a phase correction scheme in order to obtain a meaningful fidelity. In particular, we apply the operation
$R_z(-N \omega_1 \tau)$ after the DDRF sequence 
which is distinct for each nuclear spin. Note that the other contribution to the phase needs no correction since $\omega_L\tau=0$ modulo $2\pi$.  In practice, it might only be possible to track this phase classically, rather than correcting it. 
With this correction applied and the setting \(\varphi=0\), the target gate fidelity is smaller by a value on the order of $10^{-6}$ compared with setting \(\varphi\) to its optimal value. As will be seen, this is not relevant compared with the influence of other nuclear spins.

\subsection{Target nuclear spin without further spins}
\label{ssec:targetspin}

We first consider the fidelity of the target spin operation on its own.
As an example, \fref{fig:Blochsphere}~(a-b) shows the conditional evolution during a CROT gate \(R_x(\pm\tfrac{\pi}{2})\) acting on a target nuclear spin initially in the state $\ket{\ua}$, for two different values of $N$. 
As mentioned above, our default value is $N=48$, but for demonstration purposes we also include the evolution in \fref{fig:Blochsphere}~(a,b) with $N=8$, in analogy to Fig.~2 (c,d) in Ref.~\cite{Bradley2019}. Note that in the more accurate model employed here, the evolution is not idealized, i.e. it does not alternate between pure precession (azimuthal motion) and pure driving (polar motion). For this reason, to obtain a rather accurate CROT we had to choose a higher value of \(\tau\) (our standard value \(\tau = 8 \tau_L\)) which is around four times larger than in Fig.~2 (c,d) in Ref.~\cite{Bradley2019}. 
Note that since we choose $\beta=0$ for the target spin in \fref{fig:Blochsphere}~(a-d), the two rotating frames $R_0(t)$ and $R_1(t)$ introduced in App.~\ref{sec:DDRF_equations} coincide. The final states, indicated by the arrows, are drawn in the original rotating frame of the Hamiltonian \eqref{eq:Hgeneral}, after applying the azimuthal phase correction $R_z(-N \omega_1 \tau)$ introduced above.

\begin{figure}
	\includegraphics[width=\linewidth]{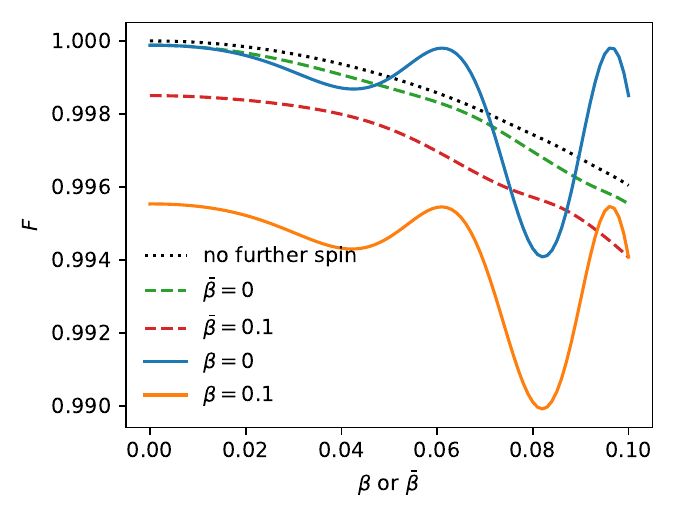}
	\caption{
    Gate fidelities of the DDRF sequence, calculated from \eref{eq:gatefidelity}, for the conditional rotation of a target nuclear spin (\(A_\parallel = 2\pi \cdot 50\ \text{kHz}\)), in the absence (black dotted line) or presence (other lines) of one unaddressed spin ($\bar{A}_\parallel = 2\pi \cdot 30\ \text{kHz}$). In the first case, the ideal evolution is given in \eref{eq:evol_(e,n)_ideal}. In the second case, the ideal evolution is given in \eref{eq:evol_ideal}, and we vary either the target parameter $\beta$ or the non-target parameter $\bar{\beta}$, keeping the other one fixed to the value indicated in the legend.
	}
	\label{fig:gatefidelity}
\end{figure}

The gate fidelity, calculated from \eref{eq:gatefidelity}, for the conditional rotation \(R_x(\pm\tfrac{\pi}{2})\), i.e.,
\begin{equation}\label{eq:evol_(e,n)_ideal}
	V_\text{ideal} = \ket{0} \bra{0} \otimes R_x(\tfrac{\pi}{2}) + \ket{1} \bra{1} \otimes R_x(-\tfrac{\pi}{2}) ,
\end{equation}
is shown in \fref{fig:gatefidelity} by the black dotted line as a function of $\beta$. 
Larger values of $N$ improve the fidelity, while larger values of $\tau$ do not (not shown). 
For the chosen parameter values (\(N=48\) and \(\tau = 8 \tau_L\)), the gate infidelity is on the order of \(10^{-6}\) for $\beta=0$.  
In a typical nuclear spin register~\cite{Bradley2019}, with target subset much smaller in number than the whole nuclear bath, a target spin can be selected with a rather small value of $\beta$, say 0.01. 
In this case, the target gate operation has a rather high fidelity and, as we will see below, the main error source is the presence of further nuclear spins, in particular, those with similar values of $A_\parallel$.
We note that in practice, the fidelity even for a single nuclear spin might be lower because of the finite duration of $\pi$-pulses and inaccurate RF frequency or phase.

\subsection{Bath spin}
\label{ssec:bathspin}

\begin{figure}
	\includegraphics[width=\linewidth]{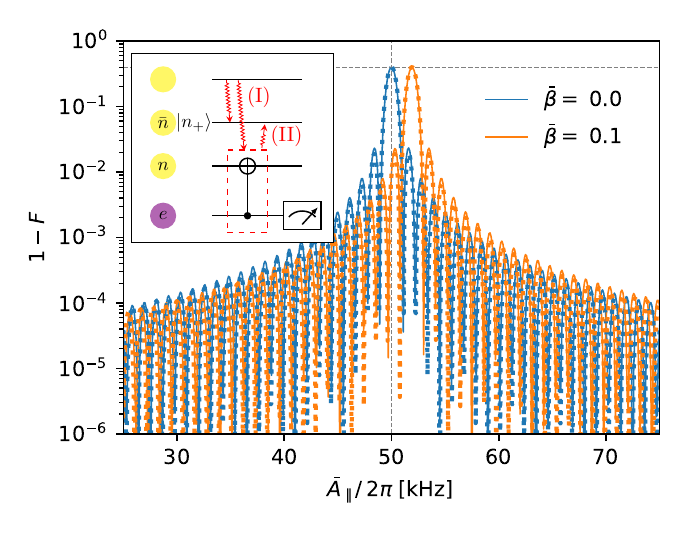}
	\caption{Infidelity of the ideally performed target CROT operation on a single target spin due to the presence of a (never addressed) bath spin. The DDRF parameters are the same as in \fref{fig:Blochsphere}~(c-d), i.e., $N=48$ , ${A_\parallel} = 2\pi \cdot 50\ \text{kHz}$, ${\beta}=0$, and $\tau=8\tau_L$. The dotted lines correspond to \eref{eq:sincsquared}. The horizontal dashed line indicates the maximal value of $1-F=0.4$.
    The inset shows the elementary process (c.f.~the dotted box in \fref{fig:circuit2nv2n}) with an additional bath nuclear spin, represented by the unlabelled yellow dot. Its decohering influence on the target spin operation (with just a single target spin or another unaddressed spin included) is represented by zigzag-shaped arrows (I) and analyzed in Sec.~\ref{ssec:bathspin}. The conditional effect of the target operation on the unaddressed spin, indicated by the arrow (II), is described by nearly the same result shown here, see Sec.~\ref{ssec:unaddrspin}.
	}
	\label{fig:bathfidelity}
\end{figure}

We now add to the system a single further nuclear spin, which is not part of the quantum register, and evaluate its influence on the gate fidelity of target nuclear spins. This decoherence effect is indicated by the arrows labeled (I) in the inset of \fref{fig:bathfidelity}.
As mentioned above, the evolution of the target spins is the same in the presence or absence of bath spins; it is just the entanglement created between the electron and the bath nucleus which negatively affects the gate fidelity. 
Following Ref.~\cite{Takou2023}, we consider $L$ nuclear spins, of which $K$ belong to the target subspace. This subspace includes both the addressed ($n$) and unaddressed ($\bar{n}$) spins (e.g., $K=2$ for the system drawn in the inset of \fref{fig:bathfidelity}), since their evolution defines our total gate fidelity. By contrast, the $L-K$ bath spins act through the quantum channel defined by the partial trace over these spins. 
The total evolution operator is written such that target spins appear before bath spins in the tensor product,
\begin{equation}
	\label{eq:Utotal}
	U = \sum_{j\in\{0,1\}} \sigma_{jj} 
	\bigotimes_{k=1}^{K} R_{\mathbf{n}_j^{(k)}} (\phi_j^{(k)})
	\bigotimes_{l=1}^{L-K} R_{\mathbf{n}_j^{(K+l)}}(\phi_j^{(K+l)}),
\end{equation}
where $\sigma_{jj} = \ket{j} \bra{j}$ are the electron's projectors, and $R_{\mathbf{n}_j} (\phi_j)$ the conditional rotation operators.
Since in this part we are just interested in the influence of the bath itself, we assume that the actual evolution (described by $U$ from \eref{eq:Utotal} with $L=K$) of the target spins is the same as the ideal one,
\begin{equation}
	\label{eq:U0target}
	U_0 = \sum_{j\in\{0,1\}} \sigma_{jj} 
	\bigotimes_{k=1}^{K} R_{\mathbf{n}_j^{(k)}} (\phi_j^{(k)}).
\end{equation}
The same choice has been made by Takou \textit{et al.}~\cite{Takou2023}, so we can follow their evaluation of the partial-trace channel of the bath spins. In the operator-sum representation \cite{Nielsen2010}, the fidelity of a general quantum operation takes the form
\begin{equation}
	\label{eq:gatefidelity_opsum}
	F = \frac{1}{d(d+1)} \sum_{i}^{} \mathrm{tr} [(U_0\hc E_i)\hc U_0\hc E_i] + | \mathrm{tr} [U_0\hc E_i] |^2,
\end{equation}
where $d=2^{K+1}$.
Here $i$ runs over the $2^{L-K}$ complete computational basis states of the environment,
and $E_i$ is the Kraus operator of the partial-trace quantum channel, which is derived from the operator $U$ in \eref{eq:Utotal} as
\begin{equation}
	\label{eq:Krausop}
	E_i = \sum_{j}^{} c_j^{(i)} p_j^{(i)} \sigma_{jj} 
	\bigotimes_{k=1}^{K} R_{\mathbf{n}_j^{(k)}} (\phi_j^{(k)}).
\end{equation}
The coefficients are defined as
\begin{align}
	c_j^{(i)} &= \prod_{m=m_1}^{m_M} \braket{\uparrow|\Psi_j^{(m)}}, \\
	p_j^{(i)} &= \prod_{s=s_1}^{s_{L-K-M}} \braket{\downarrow|\Psi_j^{(s)}},
\end{align}  
where the number $M=M(i)$ counts the zeros in the basis state $i$; when $M=0$ ($M=L-K$) then $c_j^{(i)}=1$ ($p_j^{(i)}=1$).
The state $\ket{\Psi_j^{(m)}}$ is defined analogously to the above as the final state of bath nucleus $m$ conditional on the initial electron state $j$. The initial state of all bath nuclear spins has been chosen to be $\ket{\uparrow}$ without loss of generality.
Substituting Eqs.~\eqref{eq:U0target} and \eqref{eq:Krausop} into \eref{eq:gatefidelity_opsum} and using the completeness of the Kraus operators, $\sum_{i}^{} E_i\hc E_i = \mathds{1}$, 
one finds the gate fidelity of the target subspace as
\begin{equation}
	F = \frac{1}{2^{K+1}+1} \left( 1 + 2^{K-1} \sum_{i=1}^{2^{L-K}} \Big| \sum_{j\in\{0,1\}}c_j^{(i)} p_j^{(i)}\Big|^2\right).
\end{equation} 
When the bath spins perform an unconditional rotation, $\ket{\Psi_0^{(K+l)}}=\ket{\Psi_1^{(K+l)}}$ for $l=1,\dots , L-K$, the fidelity is equal to one. This also means that the phase correction scheme applied above makes no difference here. Remember that this formula describes the influence of the bath spins on its own, since we have assumed the target evolution to be ideal. 
The formula is applicable to any number of target and bath spins. 
For $L=2$ and $K=1$ we obtain  the influence of a single bath spin on the fidelity of the single addressed nucleus (the situation shown in the inset of \fref{fig:bathfidelity} without the spin $\bar{n}$; to include spin $\bar{n}$, increment $L$ and $K$ by one),
\begin{equation}
	F = \frac{1}{5} \left( 1 + \Big| \sum_{j\in\{0,1\}}\braket{\uparrow|\Psi_j} \Big|^2 + \Big| \sum_{j\in\{0,1\}}\braket{\downarrow|\Psi_j} \Big|^2 \right).
\end{equation} 
To evaluate this scenario we run a simulation of a single DDRF sequence targeted at the addressed spin as usual, hence $F$ becomes implicitly dependant on the target spin hyperfine parameters.
Figure~\ref{fig:bathfidelity} shows the infidelity as a function of the bath spin hyperfine parameters. 
The maximal value, $1-F=0.4$, indicated by the horizontal dashed line, is obtained for orthogonal states $\ket{\Psi_0}$ and $\ket{\Psi_1}$. 
The result is well approximated by the function
\begin{equation}
	\label{eq:sincsquared}
	1-F \simeq \mathrm{sinc}^2\left(\frac{N(\bar{A}_\parallel-\bar{A}_\parallel^\mathrm{res})\tau}{2}\right), 
\end{equation}
shown by the dotted lines in \fref{fig:bathfidelity}, and 
where $\bar{A}_\parallel^\mathrm{res}$ is determined from the condition $\bar{\omega}_1=\omega_1$ for given $A_\parallel,~\beta$, and $\bar{\beta}$.
The fidelity due to the bath spins can be multiplied with our total gate fidelity from above. 
In general, our finding confirms that it is important to choose target nuclei which are spectrally well separated from unwanted nuclei. For non-vanishing bath-spin parameter $\bar{\beta}$, the choice of an integer ratio $\tau/\tau_L$ is crucial for the strong resonance feature of $1-F$, which is taken advantage of in the experiment~\cite{Bradley2019}.

\subsection{Unaddressed spin}
\label{ssec:unaddrspin}

We now inspect the evolution of an additional nuclear spin $\bar{n}$, which is not addressed during the current DDRF sequence but is neverthless part of the quantum register. The spin is supposed to be either in an initialized state ($\ket{\ua}$), or entangled with a nuclear spin in a distant node (e.g., as a result of a previous round of entanglement production and storage). In either case, the goal is to preserve its state as much as possible. 

Due to the decoupling technique, the unaddressed spin is not much affected by the driving rf pulses, and roughly performs an unconditional rotation around \(\hat{z}\), provided 
its hyperfine parameters, especially $\bar{A_\parallel}$ (where the bar over hyperfine parameters stands for an unaddressed spin), are sufficiently distinct from those of the addressed spin.
This can be seen for $\bar{\beta}=0$ and $\bar{\beta}=0.1$ in \fref{fig:Blochsphere}~(e) and (f), respectively.
The final azimuthal phase is approximately corrected by the operation $R_z(-N \omega_1 \tau)$ as described above. 
When the hyperfine parameters of the set of target spins are known, their individual acquired phases can be tracked classically 
\footnote{One would still have to figure out if tracking the phase classically would help in the case of the one-way quantum computer \cite{Raussendorf2001}.}.

With the phase correction scheme applied, the fidelity of the DDRF sequence preserving the state of unaddressed spins mainly depends on the overlap of the two final states $\ket{\Psi_{0}}$ and $\ket{\Psi_{1}}$, where the index denotes the electron state before and after the DDRF sequence. This is essentially the same effect analyzed in the previous subsection. In fact, the error probability is given by $1-|\braket{\Psi_{0}|\Psi_{1}}|^2 = 5(1-F)$, where $1-F$ is the infidelity due to a bath spin shown in \fref{fig:bathfidelity} and approximately given by \eref{eq:sincsquared}. This holds for any $\bar{A}_\parallel$ with $|\bar{A}_\parallel-\bar{A}_\parallel^\mathrm{res}|>2\pi/N\tau$, i.e., outside the region of the central peak of $1-F$. Inside this region, the error probability depends on the initial state of the unaddressed spin.
As for the  bath spin that is never addressed, the universal result outside the central peak region relies on the choice of integer ratio \(\tau/\tau_L\), at least for $\bar{\beta}\neq 0$.
We can conclude here that different target spins should also be spectrally separate.

\subsection{Combined gate fidelity}
\label{ssec:gatefid}

We are now ready to discuss the gate fidelity for the entire process creating an entanglement link, still considering only a single node in our evaluation. 
Apart from any bath spins, a simple model node (c.f.~\fref{fig:circuit2nv2n}) contains three spin qubits (electron, target nuclear spin, and unaddressed nuclear spin) with two basis states each, thus $d=2^3$ in \eref{eq:gatefidelity}.
The ideal evolution for the CROT operation in this system is
\begin{equation}\label{eq:evol_ideal}
	V_\text{ideal} = \ket{0} \bra{0} \otimes R_x(\tfrac{\pi}{2}) \otimes \sigma_0 + \ket{1} \bra{1} \otimes R_x(-\tfrac{\pi}{2}) \otimes \sigma_0.
\end{equation}
The dashed (solid) lines in Fig.~\ref{fig:gatefidelity} show the resulting gate fidelity $F$ when the tilting angle $\beta$ ($\bar\beta$) of the target (unaddressed) spin is varied. The curve labeled $\beta=0$ represents, e.g., $F$ as a function of $\bar{\beta}$ for $\beta=0$ fixed.
For $\beta=\bar{\beta}=0 $ and the chosen parameter values ($\bar{A_\parallel} = 2\pi \cdot 30\ \text{kHz}$), the infidelity $1-F$ approaches a value on the order of $10^{-4}$, which essentially corresponds to the error probability at these parameters, see \fref{fig:bathfidelity}, so it is limited by the unintentional conditional effect on the unaddressed spin. 
For increasing $\bar{\beta}$, the infidelity starts oscillating, with maximal values on the order of $10^{-2}$. This is mainly caused by the imperfection of our phase correction method, which still allows for slight rotations of the unaddressed nuclear spin from its initial state. 
The two curves $\beta=0$ and $\beta=0.1$  differ by a nearly constant factor, which is roughly the factor by which the pure target fidelity (shown by the black dotted line) decreases.
When varying $\beta$ instead of $\bar{\beta}$ (dashed lines), the overall behaviour is very similar, but the oscillations are far less pronounced. The influence of bath spins can be included according to the method described in \ref{ssec:bathspin}.

The results obtained so far can now be used to estimate a lower bound for the gate fidelity of more complex processes. For example, consider the entangling scheme of each pair of nuclear spins per node in \fref{fig:circuit2nv2n}. In each node one has to apply two instances of the elementary process highlighted by the dotted line. Considering one node, as an example, we assume that $n_1$ has $\beta_1=0$ and $n_2$ has $\beta_2=0.1$. The corresponding fidelity can be estimated from the product of two fidelity data points in \fref{fig:gatefidelity}, namely $0.996\times 0.998 \approx 0.994$. We have simulated such a scheme (for discrete parameter values only, hence no plot is shown) in order to confirm that the factorization into elementary process fidelities indeed holds. 
In the same way we can now include the other node from \fref{fig:circuit2nv2n}. Taking the example values ${F}_{ee}=0.99$ and ${F}_{en\bar{n}}=0.99$ (for each node), one obtains the fidelity for entangling $p=2$ pairs in remote nodes as
${F} = {F}_{ee}^{p} {F}_{en\bar{n}}^{2p} = 0.99^6 \approx 0.94$.
Note that our model still neglects further decoherence processes such as those caused by direct interactions between nuclear spins.

\section{Discussion}
\label{sec:discussion}

With the chosen approach of sequentially entangling remote nuclear spins, the total gate fidelity steadily declines with the number of pairs $p$. 
The concrete values strongly depends on the perpendicular hyperfine components of the target spins. Our calculated example value of 0.94 for $p=2$, even without any bath spins included, would be already rather low and might imply the necessity for entanglement purification or other correction steps. 
We have already mentioned further neglected error sources above:
finite durations of $\pi$-pulses,
decoherence due to neglected (e.g., non-secular) terms in the Hamiltonian, inaccurate RF frequency or phase, and timing issues. 
In principle, the fidelities reported in Ref.~\cite[]{Bradley2019} for entangling nuclear spins within a single node provide a better picture on what is experimentally achievable in such systems.
On the other hand, we have not included further techniques helpful for reducing error rates, in particular pulse shaping for realizing optimal control.

In the approach considered here, based on the Nemoto scheme, pairwise remote qubit entanglement is created but any intranode entanglement is avoided.
With an increasing number of pairs, more of this entanglement resource is available for fault-tolerant quantum computation.
However, the downside are gate times several orders of magnitude longer than when only the intrinsic nitrogen is used. It is questionable whether decoherence can be sufficiently avoided by DD on these time scales. In this sense the photon-based Nemoto approach with a fast electron-electron entanglement scheme seems to fit better to the intrinsic nuclear spin than to more weakly coupled nuclear spins. 
However, it might be worth discussing possible modifications to this scheme. One example would be to create multipartite intranode entanglement through the approach developed in Refs.~\cite{Takou2023,Takou2023a}. A promising solution might be the creation of one cluster (consisting of five nuclear spins) from the Nemoto approach just in one NV center. However, specific states are challenging to create by this approach for a register with random hyperfine parameters. Furthermore, the electron-nuclei gate times are still in the $\text{ms}$ regime so that decoherence remains an issue. 
%
Another use of multiple nuclear spins could be error correction within one node~\cite{Waldherr2014}.

\section{Conclusions}
\label{sec:conclusions}

We have inspected the compatibility of the photonic network approach to entangling remote NV centers with a multispin environment provided by weakly coupled $^{13}$C nuclei. In principle this is realizable by means of DD or DDRF sequences to address nuclear spins individually, for instance to initialize them or to transfer an electron-electron entanglement link onto a nuclear pair. In practice, however, some challenges  remain to be addressed. From our result for the gate fidelity we conclude that the accuracy of the multispin register strongly declines with the number of target spins, and further declines with the presence and number of unwanted spins. 
The properties of the NV center environment in diamond are  crucial for the fidelity values. 
The unconditional rotation of other target nuclear spins during a DD/DDRF sequence on one target could be another practical issue.
Realistically speaking, the use of a multispin environment with each NV center does seem to make the realization of a scalable quantum computing network easier. 

\section*{Acknowledgements}
 We acknowledge funding from the German Federal Ministry of Education and Research (BMBF) under the Grant Agreement No.~13N16212 (SPINNING).

\bibliography{draft_multispin-NV.bib}

\appendix

\section{Derivation of DDRF evolution operators}
\label{sec:DDRF_equations}

In this Appendix we recapitulate the derivation of the evolution operators describing the conditional dynamics of a nuclear spin under a DDRF sequence. The same derivation is contained in the Supplementary Material (Sec.~III.C) of Ref.~\cite[]{Bradley2019}.
The Hamiltonian \eqref{eq:Hgeneral} can be written as
\begin{equation}\begin{split}\label{eq:H_tilted}
H &= \ket{0}\bra{0}H_{0} + \ket{1}\bra{1}H_{1} , \\
H_{0} &= \omega_{L}I_{z} + 2\Omega\cos(\omega t+\phi)I_{x} , \\
H_{1} &= \omega_{1}\widetilde{I}_{z} + 2\widetilde{\Omega}_x\cos(\omega t+\phi)\widetilde{I}_{x} + 2\widetilde{\Omega}_z\cos(\omega t+\phi)\widetilde{I}_{z},
\end{split}\end{equation}
where, accounting for the tilted nuclear spin quantization axis (when the electron is in $\ket{1}$), we have introduced,
\begin{align*}
	\widetilde{I}_x &= R_y(\beta)I_xR_y(\beta)^\dagger = \cos(\beta)I_x - \sin(\beta)I_z, \\
	\widetilde{I}_y &= R_y(\beta)I_yR_y(\beta)^\dagger = I_y, \\
	\widetilde{I}_z &= R_y(\beta)I_zR_y(\beta)^\dagger = \cos(\beta)I_z + \sin(\beta)I_x, \\
	\widetilde{\Omega}_x &= \Omega\cos(\beta),\ \widetilde{\Omega}_z = \Omega\sin(\beta),
\end{align*}
and $R_y(\theta) = e^{-i\theta I_y}$.
It is convenient to define two different rotating frames depending on the electron spin state; $R_0(t) = e^{i\omega t I_z}$ and $R_{1}(t) = e^{i\omega t \widetilde{I}_z}$. 
The transformed Hamiltonians are given by
\begin{equation}\label{eq:Hint}\begin{split}
    H_{0}' &= R_{0}(t)(H_{0}-\omega I_{z})R_{0}(t)^{\dagger} \\
    &= (\omega_{L}-\omega)I_{z} + \Omega(\cos(\phi)I_{x}+\sin(\phi)I_{y}), \\
    H_{1}' &= R_{1}(t)(H_{1}-\omega \widetilde{I}_{z})R_{1}(t)^{\dagger} \\
    &= (\omega_{1}-\omega)\widetilde{I}_{z} + \widetilde{\Omega}_x(\cos(\phi)\widetilde{I}_{x}+\sin(\phi)\widetilde{I}_{y}) \\
    &= (\omega_{1}-\omega)(\cos(\beta)I_{z}+\sin(\beta)I_{x}) + \\
	&\quad\ \Omega\cos(\beta)(\cos(\phi)(\cos(\beta)I_{x}-\sin(\beta)I_{z})+\sin(\phi)I_{y}),
\end{split}\end{equation}
where the RWA has been applied.
During a DDRF sequence, the Hamiltonian alternates between $H_{0}'$ and $H_{1}'$, but since both are time-independent, the evolution can be easily calculated in a piecewise manner from $U_\sigma(t,\phi)=e^{-i H_\sigma' t}$, where $\sigma\in\{0,1\}$. However, the rotating frame needs to be changed at each electron $\pi$-pulse. This leads to the generalized evolution operators
\begin{equation}\label{eq:totalevogeneral}\begin{split}
    V_\sigma &= R_{\sigma}(2N\tau)\hc \cdot 
	U_\sigma(\tau,\phi_{N+1}) \cdot R_{\sigma}(N'\tau)\cdot \\
	&\quad\ R_{\bar{\sigma}}(N'\tau)^\dagger \cdot U_{\bar{\sigma}}(2\tau,\phi_{N}) \cdot R_{\bar{\sigma}}((N'-2)\tau) \cdot\\
	&\quad\  R_{\sigma}((N'-2)\tau)^\dagger \cdot U_{\sigma}(2\tau,\phi_{N-1}) \cdot R_{\sigma}((N'-4)\tau) \cdot \\
    &\quad\ \cdots \\
    &\quad\ R_{\sigma}(5\tau)^\dagger U_\sigma(2\tau,\phi_3)\cdot R_{\sigma}(3\tau) \cdot\\
	&\quad\  R_{\bar{\sigma}}(3\tau)^\dagger \cdot U_{\bar{\sigma}}(2\tau,\phi_2) \cdot R_{\bar{\sigma}}(\tau)\cdot \\
	&\quad\  R_{\sigma}(\tau)^\dagger \cdot U_\sigma(\tau,\phi_1), 
\end{split}\end{equation}
where $\bar{\sigma}=1-\sigma$ and $N' = 2N-1$. The final back-transformation $R_{\sigma}(2N\tau)\hc$ has been skipped in Ref.~\cite[]{Bradley2019}.

\end{document}